\documentclass{PoS}

\def\tr{{\rm Tr}}
\def\bea{\begin{eqnarray}}
\def\eea{\end{eqnarray}}

\def\msbar{\overline{\rm MS\kern-0.5pt}\kern0.5pt}

\title{The gradient flow running coupling scheme}

\ShortTitle{Running coupling}

\author{Zolt\'an Fodor\\
       University of Wuppertal, Department of Physics, Wuppertal D-42097, Germany\\
       J\"ulich Supercomputing Center, Forschungszentrum J\"ulich, J\"ulich D-52425, Germany\\
       E\"otv\"os University, Institute for Theoretical Physics, Budapest 1117, Hungary\\
       \email{fodor@bodri.elte.hu}}

\author{Kieran Holland\\
        University of the Pacific, 3601 Pacific Ave, Stockton CA 95211, USA\\
        Albert Einstein Center for Fundamental Physics, Bern University, Bern, Switzerland\\
        \email{kholland@pacific.edu}}

\author{Julius Kuti\\
        University of California, San Diego, 9500 Gilman Drive, La Jolla, CA 92093, USA\\
        \email{jkuti@ucsd.edu}}

\author{\speaker{D\'aniel N\'ogr\'adi}\\
        E\"otv\"os University Budapest, P\'azm\'any P\'eter s\'et\'any 1, 1117 Budapest, Hungary\\
        \email{nogradi@bodri.elte.hu}}

\author{Chik Him Wong\\
        University of California, San Diego, 9500 Gilman Drive, La Jolla, CA 92093, USA\\
        \email{rickywong@physics.ucsd.edu}}

\abstract{The Yang-Mills gradient flow in finite volume is used to define a running coupling scheme.
          As our main result the discrete $\beta$-function, or step scaling function, is calculated 
          for scale change $s=3/2$ at several lattice spacings for $SU(3)$ gauge theory coupled to $N_f =
          4$ fundamental massless fermions. The continuum extrapolation is performed and agreement is found
          with the continuum perturbative results for small renormalized coupling.
          The case of $SU(2)$ gauge group is briefly commented on.}

\FullConference{The 30 International Symposium on Lattice Field Theory - Lattice 2012,\\
		June 24-29, 2012\\
		Cairns, Australia}

\begin{document}

\section{Introduction}

The Yang-Mills gradient flow is a relatively new addition to the arsenal of non-perturbative tools for the study of non-abelian
gauge theories \cite{Luscher:2009eq, Luscher:2010iy, Luscher:2011bx}; 
see also \cite{Narayanan:2006rf,Lohmayer:2011si} for its use in a slightly different context. 
It has been usefully implemented for a high precision scale determination in QCD \cite{Borsanyi:2012zs} as
well as for a running coupling scheme in a finite volume setup \cite{Fodor:2012td}, among others.

All these applications involve the operator $\tr F_{\mu\nu} F_{\mu\nu}$ and its derivative with respect to
the gauge field. The discretized expression of the derivative 
is then used in a finite step integration scheme of the gradient flow,
\bea
\label{flow}
\frac{dA_\mu(t)}{dt} = - \frac{d S_{YM}(A)}{d A_\mu}\,.
\eea
The operator $\tr F_{\mu\nu} F_{\mu\nu}$ enters the simulations at two other instances. First, it is simply used in the
gauge action for generating the configurations and second in the observable\\
 $\langle E(t) \rangle = - \frac{1}{2} \langle
\tr F_{\mu\nu}(t) F_{\mu\nu}(t)$ i.e. the field strength tensor squared evaluated at flow time $t>0$. For the gauge
action we use the tree level Symanzik improved action and for $E(t)$ the same clover improved discretization is employed as
in \cite{Luscher:2009eq,Borsanyi:2012zs,Fodor:2012td}. For the derivative along the flow (\ref{flow}) we have used both
the Wilson discretization and the tree level Symanzik variant but found that the scaling properties of the latter are
more favorable for our present investigation and hence will only use that one.

Our goal is to compute the running coupling in $SU(3)$ gauge theory coupled to $N_f = 4$ flavors of 
massless fundamental fermions.
The main observable is the discrete $\beta$-function in a finite volume setup where the running is with the linear size of the
system, similarly to the Schroedinger functional \cite{Luscher:1992an}
and to other finite volume approaches. The discrete $\beta$-function corresponding 
to a scale change of $s$ is $( g^2(sL) - g^2(L) ) / \log(s^2)$, 
where the renormalized coupling $g^2(L)$ is obtained from the gradient flow.

The Schroedinger functional analysis of the same model, 4-flavor QCD, can be found in \cite{Tekin:2010mm, PerezRubio:2010ke}.

\section{Gradient flow scheme}

The gradient flow (\ref{flow}) is used to define a renormalized coupling in finite volume by first fixing the ratio
$\sqrt{8t}/L = c$ and then setting 
\bea
\label{g}
g_c^2 = \frac{128\pi^2\langle t^2 E(t) \rangle}{3(N^2-1)(1+\delta(c))}\;,
\eea
where
\bea
\label{delta}
\delta(c) = - \frac{c^4 \pi^2}{3} + \vartheta^4\left(e^{-1/c^2}\right) - 1
\eea
is given in terms of the Jacobi elliptic function $\vartheta(q) = \sum_n q^{n^2}$.
The numerical factors on the right hand side of (\ref{g}) are such that to leading order $g_{\msbar}^2 = g_c^2$ 
\cite{Fodor:2012td}. Corrections contain both even and odd powers of $g_{\msbar}$. Different choices for the constant $c$
correspond to different schemes but the leading order relationship $g_{\msbar}^2 = g_c^2$ ensures that the 1-loop
$\beta$-function in the gradient flow scheme coincides with the 1-loop $\beta$-function in the $\msbar$ scheme. The 2-loop
$\beta$-function of the $\msbar$ scheme is on the other hand not the same as in the gradient flow scheme because of the
non-zero $a_1$ term: 
$g_c^2 = g_{\msbar}^2 ( 1 + a_1(c) g_{\msbar} + O(g_{\msbar}^2) )$; for more details see \cite{Fodor:2012td}.

Massless fermions can also be included if anti-periodic boundary conditions are used. The non-trivial boundary
conditions cause the fermions to have an effective energy gap of order $1/L$ in small volume hence as one follows the running
of the coupling from the UV towards the IR at least for small renormalized coupling 
the simulation will not run into problems even at zero bare fermion mass. 

\section{A note on the $SU(2)$ case}

One might worry that for gauge group $SU(2)$ the matrix integrals in \cite{Fodor:2012td} used for the definition of the
running coupling are not finite in $D=4$ dimensions. The matrix integrals needed at leading
order are
\bea
\label{zb}
I_{D,N} = -\frac{ \int dB \frac{1}{2} \tr[B_\mu,B_\nu]^2 \exp\left( \frac{1}{2} \tr [B_\mu,B_\nu]^2 \right)}{ \int dB \exp\left(
\frac{1}{2} \tr [B_\mu,B_\nu]^2 \right) }\;,
\eea
where the integral $dB$ is over $D$ anti-hermitian traceless $N\times N$ matrices. Throughout this section we set
$L=1$. Both the numerator and denominator are finite in $D=4$ and $N>2$; see \cite{Austing:2001bd}.

Clearly, one may evaluate $I_{D,N}$ from the matrix model partition function
\bea
Z_{D,N}(b) = \int dB \exp\left( \frac{b}{2} \tr [B_\mu,B_\nu]^2 \right)
\eea
via its logarithmic derivative,
\bea
I_{D,N} = - \left.\frac{d\log Z_{D,N}(b)}{db} \right|_{b=1}\;,
\eea
provided $Z_{D,N}(b)$ and its derivative are finite. For $D=4$ and $N>2$ this is the case and one easily obtains
$I_{4,N}=N^2-1$. For $D=4$ and $N=2$ however $Z_{4,2}(b)$ is divergent. But the partition function $Z_{D,2}(b)$ can be
evaluated for $D > 4$ and is in fact finite \cite{Krauth:1998xh},
\bea
Z_{D,2}(b) = \frac{1}{2} \left(\frac{2\pi^2}{b}\right)^{\frac{3D}{4}} 
\frac{ \Gamma(D/4)\Gamma(D/4-1/2) \Gamma(D/4-1) } { \Gamma(D/2) \Gamma(D/2-1/2) \Gamma(D/2-1)}\;.
\eea
The divergence in $D=4$ is coming from the pole of $\Gamma(D/4-1)$. The logarithmic derivative of the above expression
for $D>4$ then leads to
\bea
I_{D,2} = \frac{3D}{4}\;,
\eea
which agrees with the result $N^2-1$ for $N=2$ if $D=4$ is set at the end of the calculation. Since the
perturbative calculation of $\langle t^2 E(t) \rangle$ is performed in dimensional regularization and $D=4$ is only set
at the very end, the above procedure is natural.

One needs to be very careful about corrections which are only logarithmically suppressed \footnote{DN is grateful to Martin
Luscher for an instructive discussion of this point.}. 
This question was first discussed in \cite{Coste:1985mn}.
A toy model where the issue at hand can be illustrated is
\bea
Z(g^2) = \int dx dy e^{-x^2-y^2-\frac{x^2 y^2}{g^2}}\;,
\eea
with the associated expectation value $\langle x^2 y^2 \rangle$. Let us introduce $b=1/g^2$, then
\bea
\label{int}
\langle x^2 y^2 \rangle = - \frac{ \frac{dZ(b)}{db} }{Z(b)}\;.
\eea
A key feature of the gauge theory case, a flat direction in the tree level potential, 
is shared by the above toy problem. The tree level
potential is $x^2 y^2/g^2$ and the ``1-loop'' potential $x^2 + y^2$ is subleading. For finite $g$ both the 
partition function and the expectation value are finite but the $g\to 0$ limit is quite subtle.

Naively, in the $g\to 0$ limit the ``1-loop'' potential
$x^2+y^2$ can be dropped to leading order but then both the numerator and denominator in (\ref{int}) are divergent,
similarly to the $SU(2)$ gauge theory case.

One might nevertheless first drop the $x^2+y^2$ potential and then rescale
the variables by $x\to \sqrt{g} x$ and $y\to \sqrt{g} y$ in order to obtain 
\bea
\langle x^2 y^2 \rangle = g^2 \frac{ \int dxdy x^2 y^2 e^{-x^2 y^2} }{ \int dxdy e^{-x^2 y^2} } + \ldots
\eea
where again both the numerator and the denominator are divergent. But since now the potential
is a homogeneous polynomial, the above ratio is naively $1/2$, leading to
\bea
\langle x^2 y^2 \rangle = \frac{g^2}{2} + \ldots\;,
\eea
as the naive form of the small-$g$ behavior. However exact evaluation of the integrals for finite $g$ leads to
\bea
\langle x^2 y^2 \rangle = \frac{g^2(1+g^2)}{2} - \frac{g^4}{2} \frac{K_1(g^2/2)}{K_0(g^2/2)}\;,
\eea
using the Bessel K-functions. The small-$g$ expansion of the above is
\bea
\langle x^2 y^2 \rangle = \frac{g^2}{2} \left( 1 + \frac{2}{\log(g^2/4) + \gamma_E} \right) + O(g^4)\;,
\eea
where $\gamma_E$ is Euler's constant. The naively obtained leading order $g^2/2$ result is supplemented by only
logarithmically suppressed $O(g^2/\log(g^2))$ corrections not only polynomial ones like $O(g^4)$. 

Ref. \cite{Coste:1985mn} discusses the $SU(2)$ theory, for which
a similar logarithmic correction is expected. 
Note that the coefficient of the logarithmic term for the SU(2) theory is 
at present unknown.

Since all matrix integrals to leading order for $N>2$ are finite we do not expect such complications for $SU(3)$ which
is our main application. In high orders of $g_{\msbar}$ similar logarithmic corrections may enter but
not to leading order which ensures that in the UV the 1-loop $\beta$-function is the same in our scheme as in every
other scheme.

\begin{figure}
\begin{center}
\includegraphics[width=7.0cm]{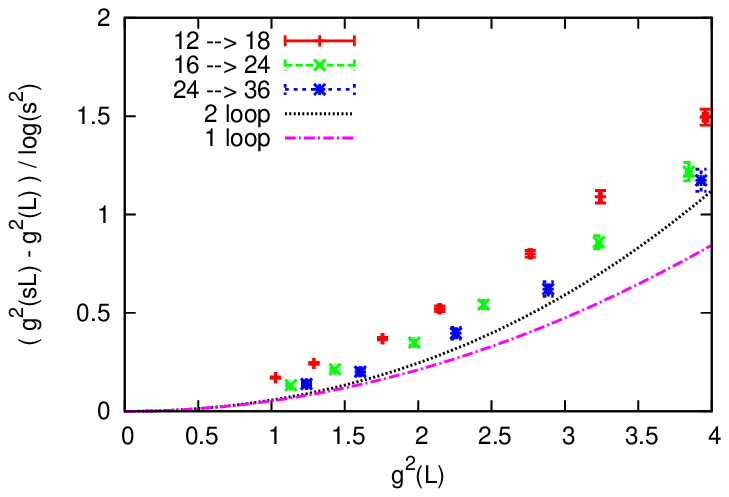} \includegraphics[width=7.9cm]{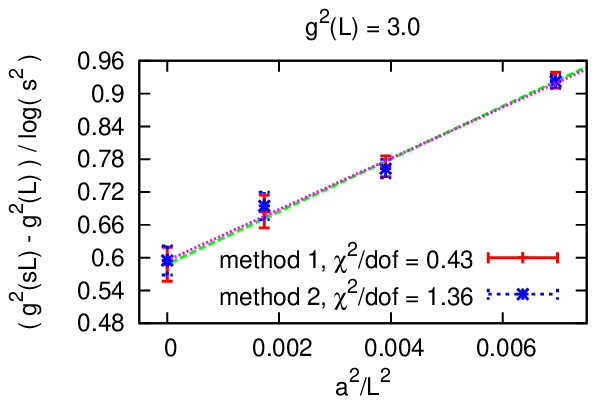}
\end{center}
\caption{Discrete $\beta$-function of $SU(3)$ gauge theory coupled to $N_f = 4$ flavors of massless fundamental
fermions. Left: scale change of
$s=3/2$, the results at 3 lattice spacings are shown together with 
the continuum universal 1-loop and $\msbar$ 2-loop results for comparison.
Right: continuum extrapolation of the discrete $\beta$-function for 
an illustrative value of the coupling $g^2(L)=3$. 
Two methods were used to perform the continuum extrapolation;
for details see \cite{Fodor:2012td}. 
The two methods give results which agree with each 
other for all of our $g^2(L)$ values.}
\label{beta}
\end{figure}

\section{Results}

Once the renormalized coupling is defined by (\ref{g}) the discrete $\beta$-function, or step scaling
function \cite{Luscher:1991wu}, can be computed by increasing the volume by a factor of $s$. The simulations were performed with the same parameters as reported in \cite{Fodor:2012td}.

\begin{figure}
\begin{center}
\includegraphics[width=9cm]{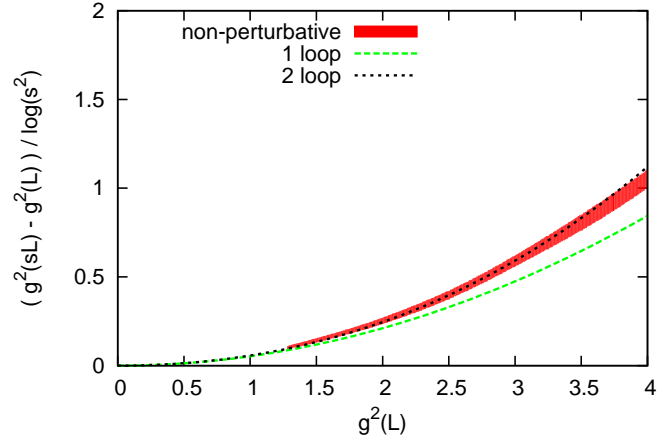}
\end{center}
\caption{Discrete $\beta$-function of $SU(3)$ gauge theory coupled to $N_f = 4$ flavors of massless fundamental fermions for a
scale change of $s=3/2$. The continuum extrapolated result is shown together with the universal 1-loop and $\msbar$ 2-loop 
results for comparison \cite{Fodor:2012td}.}
\label{cont}
\end{figure}

For the scale change $s=3/2$ we have results for 4 lattice spacings, corresponding to the volume changes 
$8\to 12$, $12\to18$, $16\to24$ and $24\to36$. 
The coarse lattice spacing, $8\to12$, 
is definitely outside of the $a^2$ scaling region and can not be 
used for continuum extrapolation. Thus we do not discuss 
these coarse lattices any further. 
The results are shown in the left
panel of figure \ref{beta} with the choice of $c=0.3$.
(Note that as a cross check we repeated the calculations with 
the scale change $s=2$ and with 3 lattice spacings,
corresponding to $8\to16$, $12\to24$ and $18\to36$. All the findings are
similar to those obtained with $s=3/2$.)

For the $s=3/2$ case with 3 lattice spacings and volume changes
$12\to18$, $16\to24$ and $24\to36$ the results can be
continuum extrapolated assuming a fit linear in $a^2/L^2$; see the
right panel of figure \ref{beta} and figure \ref{cont}.

\section{Conclusions}

In this work a running coupling scheme was investigated where the running scale $\mu$ is given by the linear size of the system
$\mu = 1/L$. The main idea is to use the Yang-Mills gradient flow 
\cite{Luscher:2009eq,Luscher:2010iy,Luscher:2011bx} but
to adapt it to a finite volume setting \cite{Fodor:2012td}. 
In principle the original infinite volume construction can also be used to define a scheme
where the running is via $\mu = 1/\sqrt{8t}$ where $t$ is the flow time, however in this setup one would need to ensure
that finite volume effects are fully under control.
In contrast, incorporating finite volume dependence explicitly into the setup eliminates this problem.
In addition, a small finite physical volume with appropriate boundary conditions for the fermions guarantees a gap in
the spectrum even in the massless case. 

Our goal was to determine $g^2(sL)$ as a function of  $g^2(L)$ at various
lattice spacings. To 
that end we measured the renormalized coupling $g^2(L)$ and $g^2(sL)$
with the same set of bare couplings, $m=0$ and $\beta$. In the next step we 
performed an extrapolation to the continuum limit, $a^2/L^2 \rightarrow 0$,
for the discrete $\beta$-function.

For our main results we used simulations at 3 lattice spacings 
with $s=3/2$ and $c=0.3$. 
Renormalized couplings between $0$ and $2.0$ were studied and
a controlled continuum extrapolation was carried out.
For small renormalized coupling the universal continuum 1-loop result is reproduced and the
$\msbar$ 2-loop result is also consistent with our numerical results; see figure \ref{cont}.
This suggests that for our choice of $s=3/2$ and $c=0.3$
the coefficient $a_1$  
connecting the gradient flow scheme to the $\msbar$ scheme $g_c^2 = g_{\msbar}^2 ( 1 + a_1(c) g_{\msbar} + \ldots)$ is
probably small.

\section*{Acknowledgments}

The work presented here was supported by the DOE under grant DE-FG02-90ER40546, by the NSF under
grants 0704171 and 0970137, by the EU Framework Programme 7 grant (FP7/2007-2013)/ERC
No 208740, and by the Deutsche Forschungsgemeinschaft grant SFB-TR 55. 
Computations were performed on the GPU clusters at the Eotvos University in Budapest, Hungary and the 
University of Wuppertal, Germany using the CUDA port of the code \cite{Egri:2006zm}. Kalman Szabo and Sandor Katz are
gratefully acknowledged for code development. KH wishes to thank the Institute for Theoretical Physics and the Albert Einstein
Center for Fundamental Physics at Bern University for their support. DN would like to thank Kalman Szabo for
suggesting to look into \cite{Luscher:2010iy} for possible applications and 
especially Pierre van Baal for very useful discussions.
KH and JK wish to thank the Galileo Galilei Institute for Theoretical
Physics and INFN for their hospitality and support at the workshop
"New Frontiers in Lattice Gauge Theories". We also wish to thank Martin
Luscher for discussions on special properties of SU(2) integrals at 
the workshop and his instructive correspondence with DN on this topic.

\end{document}